\documentclass[aps,prb,floatfix,amsmath,amssymb,superscriptaddress,twocolumn,nofootinbib]{revtex4-2}
\usepackage{bm}                     
\usepackage{appendix}
\usepackage[latin1]{inputenc}
\usepackage{dcolumn}      
\usepackage{bbm}
\usepackage{color}
\usepackage{xcolor}

\usepackage[mathscr]{eucal}
\usepackage{latexsym}
\usepackage{amsbsy}
\usepackage{float}
\usepackage{graphicx}
\usepackage{epsfig}
\usepackage{subfigure}
\usepackage{wrapfig}
\usepackage{epsf}
\usepackage{float}
\usepackage[normalem]{ulem}
\usepackage{qcircuit}

\definecolor{darkblue}{HTML}{004D6B}
\definecolor{darkred}{HTML}{8c1515}
\definecolor{darkgreen}{HTML}{006400}
\usepackage{hyperref}
\hypersetup{ colorlinks=true, urlcolor=darkblue, citecolor=darkred,
    linkcolor=darkblue, breaklinks }
\usepackage{parskip}

\newcommand{\be}{\begin{equation}}
\newcommand{\ee}{\end{equation}}
\newcommand{\ba}{\begin{array}{l}}
\newcommand{\ea}{\end{array}}
\newcommand{\re}[1]{(\ref{#1})}
\newcommand{\ci}[1]{\cite{#1}}
\newcommand{\banonum}{\begin{eqnarray*}}
\newcommand{\eanonum}{\end{eqnarray*}}
\newcommand{\baa}{\begin{eqnarray}}
\newcommand{\eaa}{\end{eqnarray}}
\newcommand{\bfr}{\begin{flushright}}
\newcommand{\efr}{\end{flushright}}
\newcommand{\bfl}{\begin{flushleft}}
\newcommand{\efl}{\end{flushleft}}
\newcommand{\lab}[1]{\label{#1}}

\begin{document}

\title{Coulomb impurities in graphene driven by ultrashort electromagnetic pulses: Excitation, ionization, and pair creation}

\author{Saparboy Rakhmanov}
\affiliation{Chirchik State Pedagogical University, 104 Amur Temur Str., 111700 Chirchik, Uzbekistan}
\author{Reinhold Egger}
\affiliation{Institut f\"ur Theoretische Physik, Heinrich-Heine-Universit\"at, 40225  D\"usseldorf, Germany}
\author{Davron Matrasulov}
\affiliation{Turin Polytechnic University in Tashkent, 17 Niyazov Str., 100095 Tashkent, Uzbekistan}

\begin{abstract}
We provide a theory for electronic transitions induced by ultrashort electromagnetic pulses
in two-dimensional artificial relativistic atoms which
are created by a charged impurity in a gapped graphene monolayer. 
Using a non-perturbative sudden-perturbation approximation, we derive and discuss
analytical expressions for the probabilities for excitation, ionization and electron-hole pair creation in this system.
\end{abstract}
\maketitle

\section{Introduction}\label{sec1}

Graphene monolayers represent a unique two-dimensional (2D) material with a 
broad spectrum of unusual and practically important properties \cite{CastroNeto}. 
Apart from its application potential, a remarkable feature of graphene
is that it provides a table-top laboratory for testing relativistic 
quantum physics in two spatial dimensions under strong-coupling conditions.
This feature is related to the fact that the planar quantum electrodynamics (QED) 
in graphene is characterized by a coupling constant (the effective fine structure constant) that can be much larger than in conventional three-dimensional 
QED \cite{Kotov,Pereira,Dipole}. To elucidate the resulting strongly coupled 
2D relativistic quantum physics, one may create and study
artificial relativistic atoms by doping charged impurities into monolayer 
graphene \cite{Luican,Wang,Wang2,Pereira,Gamayun,Dipole,Denis}.

The electron dynamics in 2D graphene monolayers effectively operates under relativistic conditions
similar to the regime realized by high-$Z$ atoms in ordinary QED. 
This correspondence allows one to observe different supercritical effects \cite{Kotov},
such as vacuum polarization, pair creation, or Klein tunneling, 
which occur when the atomic electron's energy level reaches the boundary of the Dirac sea. 
Earlier, such effects were extensively studied in the context of conventional 3D QED \cite{Greiner1,Popov1}. However, their experimental realization is still considered 
a complicated and expensive task. On the other hand, recent experiments on
Coulomb impurities in graphene have already shown manifestations of related phenomena  \cite{Luican,Wang,Wang2}. 

In this paper, we study electronic transitions for relativistic 2D atoms in gapped graphene monolayers 
induced by the interaction of the atom with an ultrashort electromagnetic pulse. The external
pulse can be created either by picosecond (or shorter)  lasers, or by a fast and highly charged ion passing near the surface of the graphene sheet. Our motivation for this study comes from the fact that the experimental realization of high-$Z$ atoms or ions interacting 
with strong electromagnetic fields represents a highly challenging task which
requires facilities with high-power lasers or accelerators for highly charged ions \cite{Greiner1,Popov1,Baltz01,Baltz02,Baltz03,Eichler,Thomas2015,Hag2013,Hill2013,Kovalenko}. Using graphene for such studies may provide an alternative low-cost table-top realization. This concerns 
especially supercritical phenomena such as particle-antiparticle pair creation and vacuum polarization \cite{Kotov,Novikov,Allor,Lewk,Lewk1,Most1,Fillion,Akal01,Akal02,Golub,selym}, whose experimental study otherwise requires large-scale facilities 
for stripping and accelerating high-$Z$ atoms and highly charged ions, e.g.,
LHC (CERN), FAIR (GSI), or RHIC (BNL).  
Since electronic transitions of the atom due to its interaction with an ultrashort electromagnetic pulses cannot be calculated within perturbation theory (e.g., using the Born approximation), 
one needs to apply non-perturbative methods. One such approach has been developed 
by Matveev in Ref.~\cite{VI}, see also Refs.~\cite{VI2009,VI2016}.  Here we use Matveev's
approach to calculate probabilities and differential cross sections for the excitation, ionization and particle-antiparticle pair creation in a planar artificial
atom graphene. 
We note that particle-antiparticle pair creation may be viewed 
as ``ionization" of the Dirac sea, where an electron from the
filled Dirac sea is excited to the upper continuum. 
Unlike in 3D QED with high-$Z$ atoms, such transitions are easier to realize
in graphene with a Coulomb impurity and one does not need to employ high-intensity laser fields. 
Electron-hole pair creation via the Schwinger mechanism in graphene with external time-dependent electromagnetic fields has previously been studied in Refs.~\cite{Lewk,Akal01,Akal02,selym}.

The remainder of this paper is organized as follows. In Sec.~\ref{sec2}, mainly following Ref.~\cite{Novikov}, we briefly 
summarize the electronic states of planar relativistic atoms in gapped graphene monolayers, which represents a relativistic 2D Coulomb problem. 
Section \ref{sec3} presents a detailed description of the sudden-perturbation approximation for this artificial relativistic atom interacting with an ultrashort electromagnetic pulse. 
In Sec.~\ref{sec4}, we present results for the
probabilities and cross sections of different electronic transitions, including the case of 
particle-antiparticle pair creation. Finally, Sec.~\ref{sec5} provides 
concluding remarks. Throughout, we use units with $\hbar=e=1$.

\section{Relativistic atoms in graphene}\label{sec2}

We here briefly summarize the electronic structure of 2D artificial relativistic atoms
in graphene, mainly following Ref.~\cite{Novikov}.
Doping a charged Coulomb impurity into a gapped graphene monolayer causes a capture of free electrons which form a 2D artificial atom. The electron dynamics of such an atom is described in terms of the 
2D relativistic Dirac equation \cite{Novikov,Khalilov}. 
Basic mechanisms for vacuum polarization effects, including quasiparticle pair creation in graphene, induced by time-dependent electromagnetic fields have been studied in Refs.~\cite{Kotov,Novikov,Khalilov,Allor,Lewk,Lewk1,Most1,Fillion,Akal01,Akal02,Golub,selym}.
In particular, the critical charge for supercriticality, at which the lowest bound-state energy level dives
into the filled Dirac sea, turns out to be much lower than for 3D atoms \cite{Novikov,Khalilov}.

In the presence of a homogeneous quasiparticle energy gap $M$, which may arise, e.g., due to strain effects or due to a substrate superlattice potential \cite{CastroNeto}, the single-electron dynamics in a graphene monolayer with a Coulomb impurity at the origin is
described by the 2D Dirac Hamiltonian \cite{CastroNeto,Novikov,Kotov},
\be
H_0 = v_F(\sigma_x p_x +\sigma_y p_y)+M\sigma_z -\frac{v_F\alpha}{r} \sigma_0 .
\lab{hamilt}
\ee 
Here $v_F$ is the Fermi velocity, and the dimensionless number $\alpha$ 
parametrizes the charge of the Coulomb impurity (multiplied by the 
fine structure constant and the inverse dielectric constant). We use ${\bf r}=(x,y)$, $r=\sqrt{x^2+y^2}$, $p_{x,y} =-i\partial_{x,y}$, and the standard Pauli matrices
$\sigma_{x,y,z}$ with the identity $\sigma_0$.  
These matrices act in the sublattice space of graphene's honeycomb lattice \cite{CastroNeto}, where we focus on a single K valley in the band structure. 
We thus assume that the electromagnetic field of the pulse varies smoothly in space
on the scale of the lattice constant ($a_0\sim 2.46$~\AA) \cite{CastroNeto}. 
In addition, we assume that no spin-dependent physics is of importance such that the electron spin can be kept implicit.

In the absence of external fields, the quasiparticle 
spinor wave function obeys the stationary Dirac equation, $H_0|\psi\rangle =E|\psi\rangle$.
Following Refs.~\cite{Novikov,Khalilov}, we separate the angular ($\phi$) and radial ($r$) variables as
\be \label{FG}
\psi_{n,j}({\bf r})=\left(\begin{array}{c}F_{n,j}(r) \,\Phi_{j-1/2}(\phi)\\
iG_{n,j}(r)\, \Phi_{j+1/2}(\phi)\end{array}\right),
\ee
where the integers $n$ and half-integers $j$ are principal and angular quantum numbers, respectively.
With integer $m$,  the angular eigenfunctions are given by $\Phi_m(\phi)=\frac{1}{\sqrt{2\pi}}e^{im\phi}$. 
The radial eigenfunctions $F_{n,j}(r)$ and $G_{n,j}(r)$ are specified in closed form in Ref.~\cite{Novikov}.
For the discrete spectrum with energies $|E|<M$, and assuming $0\le \alpha<1/2$ throughout,
the eigenenergies are given by 
\be\label{eigenen}
E_{n,j} =\frac{M}{\sqrt{1+\frac{\alpha^2}{(n+\gamma_j)^2}}},\quad \gamma_j =\sqrt{j^2 -\alpha^2},
\ee
where $n\ge 0$ for $j>0$, and $n>0$ for $j<0$.

Let us next provide explicit expressions for planar spinor waves of a free Dirac particle and for 
scattering states in the  Coulomb field \cite{Novikov}.
These wave functions are employed below for a calculation of the pair creation and ionization probabilities caused by ultrashort electromagnetic pulses. 
Planar wave states with momentum ${\bf p}$ and
energy dispersion $E=\pm\sqrt{v_F^2p^2+M^2}$ (with $p=|{\bf p}|$) can be written as
\be
\psi_{\pm|E|;\textbf{p}} =  \frac{e^{i\textbf{p}\cdot {\bf r}}}{\sqrt{2|E|}}\left(
\begin{matrix}
    \sqrt{|E+M|}  \\
    \pm\sqrt{|E-M|}e^{i\phi_\textbf{p}}  \\
    \end{matrix}
\right), 
\ee
where $\phi_{\bf p}=\arg{(p_x+ip_y)}$.
Similarly, spinor spherical waves are given by
\be 
\psi_{p,j}(\textbf{r})=\frac{1}{\sqrt{2|E|}}\left(
\begin{matrix}
    \sqrt{|E+M|}\, R_{p,j-1/2}(r)\,\Phi_{j-1/2}(\phi)  \\
    \pm i\sqrt{|E-M|}\, R_{p,j+1/2}(r)\,\Phi_{j+1/2}(\phi) \\
    \end{matrix}
\right), \ee
where $R_{p,m}(r)=\sqrt{2\pi p}\,J_m(pr)$  with Bessel functions $J_m$.

Finally, we turn to scattering states in the Coulomb field, with
energies in the continuous part of the spectrum $(|E|>M$). In the asymptotic 
regime $pr\to \infty$, they are given by Eq.~\eqref{FG} with
\begin{eqnarray}\label{scat}
F_{p,j}&\simeq&\sqrt{\frac{2|E+M|}{|E|r}}\cos(pr-j\pi/2+\alpha_E\ln
2pr+\delta_j), \\ \nonumber
G_{p,j}&\simeq&\pm\sqrt{\frac{2|E-M|}{|E|r}}\sin(pr-j\pi/2+\alpha_E\ln
2pr+\delta_j),
\end{eqnarray}
where we use
\begin{eqnarray} 
e^{-2i\xi_j}&=&\frac{\gamma_j-i\alpha_E}{j+iM\alpha/v_Fp},  \quad \alpha_E=\frac{\alpha E}{v_Fp},
\\  \nonumber
\delta_j&=&\xi_j+\frac{\pi}{2}(j-\gamma_j)-\arg\Gamma\bigg( 1+\gamma_j+\frac{i\alpha E}{v_Fp} \bigg),
\end{eqnarray}
with the Gamma function $\Gamma$.
Below, the spinor wave function in Eqs.~\eqref{FG} and \eqref{scat} is used 
as final state $|\psi_f\rangle$ 
in our calculations of the ionization probabilities and cross sections. 

\section{Sudden-perturbation approximation}\label{sec3}

The sudden-perturbation approximation is a powerful tool for describing atoms in external time-dependent fields if the duration of the external pulse is much shorter than the 
characteristic period of the electron bound to the first Bohr orbit of the atom.
Here we will adopt the sudden-perturbation approach of Ref.~\cite{VI} to  
an artificial atom in a 2D graphene layer, see Sec.~\ref{sec2}, 
in the presence of an ultrashort electromagnetic pulse.

In the presence of a spatially inhomogeneous ultrashort pulse $U({\bf r},t)$, the spinor wave function dynamics is governed by the 2D time-dependent Dirac equation,
\be
i\frac{\partial \Psi({\bf r},t)}{\partial t} = [H_0 + U({\bf r},t)]\Psi({\bf r},t),
\lab{Dir3}
\ee
with $H_0$ in Eq.~\re{hamilt}. Describing the electromagnetic pulse
in terms of a vector potential ${\bf A}({\bf r},t)$ and
a scalar potential $\varphi({\bf r},t)$, we have
\be
U({\bf r}, t) =\frac{v_F}{c}{\bm\sigma}\cdot{\bf A} -\varphi \sigma_0,
\ee
where ${\bm\sigma}=(\sigma_x,\sigma_y)$. We thus obtain 
\be
i\dot\Psi= \left[v_F{\bm\sigma}\cdot\left({\bf p}+\frac{1}{c}{\bf A} \right) 
+M\sigma_z- \left(\varphi +\frac{v_F\alpha}{r}\right)\sigma_0\right]\Psi.
\lab{pert1}
\ee
We now assume a gauge where the scalar potential vanishes,  $\varphi=0$,  and the
vector potential ${\bf A}({\bf r},t)={\bf A}(\eta({\bf r},t))$ only depends on $({\bf r},t)$ through
the variable 
\be
\eta({\bf r},t) =\omega_0 t -{\bf  k}_0\cdot {\bf r} ,\quad \omega_0=c|{\bf k}_0|.
\ee
The pulse maximum is then described by the wave vector ${\bf k}_0$ with frequency $\omega_0$.
Next we use the gauge transformation \ci{VI}
\be 
 {\bf A}'={\bf A}+\nabla f, \quad
\varphi'=\varphi-\frac{1}{c}\frac{\partial f}{\partial t},  \quad f=-{\bf  A}\cdot {\bf r}. \lab{trans1}
\ee 
We thereby obtain
\begin{eqnarray}
    {\bf A}'&=&\left({\bf r} \cdot \frac{d{\bf A}}{d\eta}\right){\bf k}_0,\\ 
    \nonumber
\varphi'&=&\left( {\bf r}\cdot \frac{d {\bf A}}{d\eta} \right)|{\bf k}_0|=-{\bf E}\cdot {\bf r},
\end{eqnarray}
with the electric field ${\bf E}=-|{\bf k}_0| \frac{d {\bf A}}{d\eta}.$

\begin{figure}[t!]
\centering
\includegraphics[totalheight=0.27\textheight]{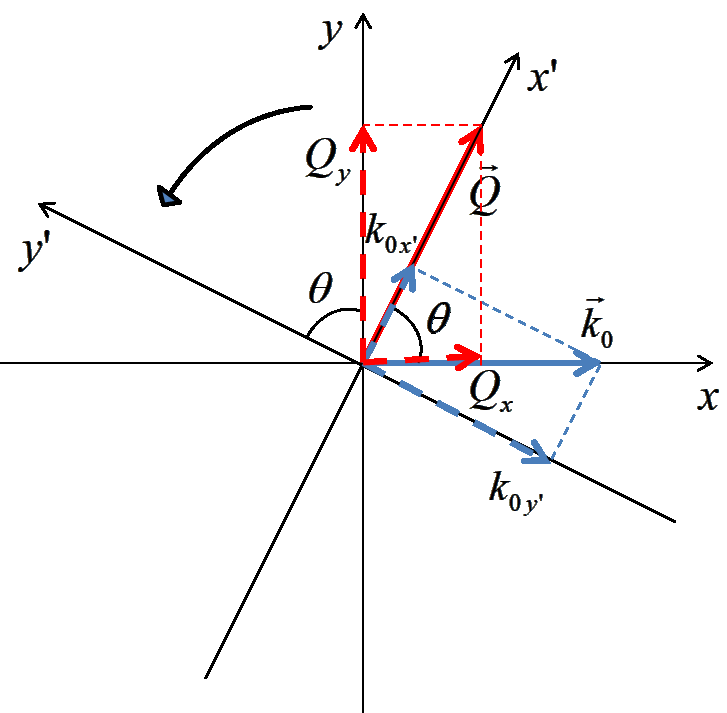}
 \caption{Rotation of the coordinate system.} \label{fig1}
\end{figure}

To proceed, let us assume that the wave vector ${\bf k}_0$ is directed along the $x$-axis, 
see Fig.~\ref{fig1}. 
In the new gauge, the external electromagnetic field can then be written as
\begin{eqnarray}\nonumber
U({\bf r},t)&=&\frac{v_F}{c}{\bm \sigma}\cdot{\bf A}'-\varphi'\sigma_0=
-\left(\sigma_0-\frac{v_F}{c}\frac{{\bm \sigma}\cdot {\bf k}_0}{|{\bf k}_0|}\right)\varphi'=\\
&=& -\left(\sigma_0-\frac{v_F}{c}\sigma_x\right)\varphi'.
\lab{potent}
\end{eqnarray}
The evolution of the spinor wave function from time $t_0$ to time $t$ is then governed by
\be
|\Psi(t)\rangle ={\cal T}\exp\left(-i\int_{t_0}^{t}dt' U(t')\right)|\Psi(t_0)\rangle,
\lab{WF0}
\ee
where ${\cal T}$ is the time ordering operator.
For ultrashort pulses, one finds $[U(t),U(t')]=0$ for all time pairs, 
see also Ref.~\cite{VI}, and hence
the time ordering operator can effectively be omitted in Eq.~\eqref{WF0}. 
We may thus calculate the transition amplitude $a_{f,i}$ from an initial state $|\psi_i\rangle$ 
to a final state $|\psi_f\rangle$ by using the relation 
\be\lab{proba02}
a_{f,i} = \langle\psi_f|\exp\left(-i\int_{-\infty}^{+\infty}dt U(t) \right)|\psi_i\rangle. 
\ee
According to Ref.~\cite{VI}, the transition amplitude in a driven system described by Eq.~\re{Dir3} is given by Eq.~\re{proba02} if the perturbation has a $\delta$-function like time dependence, 
\be
\tilde{U}(t)=U_0\delta(ct-x), \quad U_0=c\int_{-\infty}^{+\infty}dt\, U(ct-x). \lab{deltapot}
\ee
In this limit, the sudden-perturbation approach becomes exact.  However, even if
the $\delta$-function limit is only realized approximately, the obtained relations 
can be used for accurately solving the time-dependent Dirac equation \re{Dir3}.

In what follows, we work in the new gauge given by Eq.~\re{trans1},  where we omit the primes henceforth. 
Substituting the potential \re{deltapot} into Eq.~\re{Dir3}, we find 
\be
i\dot\Psi = \left[v_F{\bm \sigma}\cdot{\bf p} + M\sigma_z -\frac{v_F\alpha}{r}\sigma_0-\left(\sigma_0-\frac{v_F}{c}\sigma_x\right)\tilde\varphi\right]\Psi,
\lab{pert2}
\ee
where the ultrashort pulse is described by
\be
\tilde\varphi = \varphi_0\delta(ct-x),\quad \varphi_0 = c\int_{-\infty}^{+\infty}dt\,\varphi(ct-x).
\lab{pulse1}
\ee

The solution of Eq.~\re{pert2} can be expanded in the complete set of eigenfunctions of the unperturbed system, see Sec.~\ref{sec2}. With the 2D spinor wave functions $\psi_l({\bf r})$ in Eq.~\eqref{FG} and the corresponding eigenenergies $E_l$, where $l=(n,j)$ combines principal ($n$) and angular momentum ($j$) quantum numbers, we write
\be
\Psi({\bf r}, t) = \sum_l a_l(t) e^{-iE_lt}  \psi_l({\bf r} ),
\lab{wf00}
\ee
with complex-valued and time-dependent expansion coefficients $a_l(t)$.
Substituting Eq.~\re{wf00} into Eq.~\re{pert2}, the time dependence 
of the coefficient $a_f(t)$ for the final state $|\psi_f\rangle$ is governed by
\begin{equation}
\frac {da_{f}(t)}{dt}=-ie^{iE_f t}\langle\psi_f|\tilde
U(t)|\Psi_i\rangle,\label{ampl00}
\end{equation}
where we assume that prior to the interaction, the atomic electron was
in the state $|\psi_i\rangle$:
\begin{equation}
\Psi_i({\bf r},t)|_{t \to -\infty}= e^{-iE_it} \psi_i({\bf r}),
\label{eq:a2}
\end{equation}
with the initial condition
\begin{equation}
a_{f}(t)|_{t\to -\infty}= \delta_{f,i}.\label{eq:a3}
\end{equation}

For $\tilde U(t)=U_0\delta(ct-x)$, in order to solve Eq.~\re{ampl00}, it is sufficient to
determine $\Psi({\bf r},t)$ only near $x=ct$ \cite{VI}. This can be implemented by transforming to 
light-cone coordinates,
\begin{equation}
x^-=ct-x, \quad x^+=ct+x,
\label{eq:a4}
\end{equation}
where time and space derivatives of the wave function can be written as
\begin{eqnarray} 
\dot{\Psi}&=&\frac{\partial\Psi}{\partial x^-}\frac{\partial x^-}{\partial t}=c\frac{\partial\Psi}{\partial x^-},\\ \nonumber
\frac{\partial\Psi}{\partial x}&=&\frac{\partial\Psi}{\partial x^-}\frac{\partial x^-}{\partial x}=-\frac{\partial\Psi}{\partial x^-}.
\end{eqnarray}
In the immediate vicinity of $x^-=0$,  Eq.~\re{pert2} can be written as \ci{VI}
\begin{equation}
ic\left(\sigma_0-\frac{v_F}{c}\sigma_x\right)\frac {\partial \Psi}{\partial x^-}=
-\left(\sigma_0-\frac{v_F}{c}\sigma_x\right)\tilde\varphi\,\Psi,\label{eq:a5}
\end{equation}
with $\tilde{\varphi}=\varphi_0\delta(x^-)$.
With the Heaviside step function $\theta(x)$, where we use
$\frac{d}{dx}\exp\theta(x)=\delta(x)\exp\theta(x)$,
the solution of Eq.~\re{eq:a5} can be written as 
\be 
\Psi(x^-+\varepsilon)=\exp\left[i\frac{\varphi_0}{c}\theta(x^-)\right]\Psi(x^--\varepsilon),
\lab{sol00}
\ee
where $\varepsilon$ is an infinitesimal positive quantity.
In terms of normal time and space coordinates with $t = x/c + \varepsilon$, Eq.~\re{sol00} 
takes the form
\be 
\Psi({\bf r},t)=\exp\left[i\frac{\varphi_0}{c}\theta(ct-x)\right]\,e^{-iE_it}\,\psi_i({\bf r}). \lab{solut06}
\ee
Next we substitute Eq.~\re{solut06} into Eq.~\re{ampl00} and integrate over time. Taking into account  the initial condition \re{eq:a3}, we obtain the transition amplitude $a_{f,i}$ in 
Eq.~\eqref{proba02} as
\begin{eqnarray}\label{ampl00001} 
&& a_{f,i} = a_f(t\to \infty) = \delta_{f,i}+i\int_{-\infty}^{+\infty}dt\,
e^{i(E_f-E_i)t} \times  \\ &&\times 
 \langle\psi_f|\left(\sigma_0-\frac{v_F}{c}\sigma_x\right)\varphi_0\exp\left[i\theta(ct-x)\frac{\varphi_0}{c}\right] \delta(ct-x)|\psi_i\rangle.
\nonumber
\end{eqnarray}
Performing the integral, we obtain
\begin{eqnarray}\nonumber
a_{f,i}&=&\delta_{fi}+
\langle\psi_f|\left(\sigma_0-\frac{v_F}{c}\sigma_x\right) \times \\
&\times& e^{i(E_f-E_i)x/c} \left(e^{i\varphi_0/c}-1\right)|\psi_i\rangle.
\label{ampl00002}
\end{eqnarray}
Using the relation
\be 
\langle\psi_f|\left(\sigma_0-\frac{v_F}{c}\sigma_x\right)|\psi_i\rangle=\delta_{f,i},
\label{relation0003} 
\ee
we finally arrive at the transition amplitude
\begin{equation}
a_{f,i}=\langle\psi_f|\left(\sigma_0-\frac{v_F}{c}\sigma_x\right)\, 
e^{i(E_f-E_i)x/c}\, e^{i\varphi_0/c} |\psi_i\rangle.
\label{ampl1}
\end{equation}
Using Eq.~\re{ampl1}, we can now calculate transition amplitudes between electronic states of the artifical atom corresponding to excitation, ionization, or electron-hole pair creation processes.

We choose the external electromagnetic field as a Gaussian pulse 
of short time duration $\lambda$,
\begin{equation}
{\bf E}({\bf r},t)={\bf E}_0\exp\left[-\frac{1}{\lambda^2}\left(t-\frac{{\bf k}_0\cdot{\bf r}}{\omega_0}\right)^2\right]\cos(\omega_0t-{\bf k}_0\cdot {\bf r}).\label{eq:13}
\end{equation}
Approximating this pulse shape by a $\delta$-potential as in Eq.~\re{pulse1}, we 
find $\tilde \varphi=-{\bf r}\cdot {\bf E}({\bf r},t)$ and therefore
\be
 \varphi_0=c\int_{-\infty}^{+\infty}dt\, \tilde\varphi(ct-x) =c {\bf q}\cdot {\bf r},
\ee
where we define
\begin{equation}
{\bf q}=-\int_{-\infty}^{+\infty}dt\, {\bf E}({\bf r},t) =
-\frac{\sqrt{\pi}}{\lambda}\, e^{-\omega_{0}^{2}/(2\lambda)^2}\,{\bf E}_0 .\label{eq:a14}
\end{equation}
The quantity ${\bf q}$ describes the momentum transfer due to the 
interaction of external electromagnetic pulse with the atom.
The dependence of transition probabilities and cross sections on ${\bf q}$ can be probed experimentally.

In order to simplify the following calculations, we perform a rotation of the 
coordinate system as illustrated in Fig.~\ref{fig1}.  
Taking into account that $\sigma_x = {\bm \sigma}\cdot {\bf k}_0/k_0$,
and introducing the vector ${\bf Q}=(Q_x,Q_y)=(\Omega_{f,i}/c,q)$, with the transition frequency $\Omega_{f,i}=E_f-E_i$ and $q=|{\bf q}|$, 
the transition amplitude in Eq.~\re{ampl1} can be rewritten as
\begin{equation}
a_{f,i}=\langle\psi_f|\left(\sigma_0-\frac{v_F}{c}\frac{{\bm \sigma}\cdot {\bf k}_0}{k_0}\right)
 e^{i{\bf Q}\cdot {\bf r}}|\psi_i\rangle.\label{eq:a15}
\end{equation}
In the rotated system, the $x'$-axis is aligned with ${\bf Q}$, see Fig.~\ref{fig1}. 
In what follows, we omit the primes again and 
define the angle $\theta$ through 
\be
\frac{{\bm\sigma}\cdot {\bf k}_0}{k_0}=\sigma_x\cos\theta-\sigma_y\sin\theta .\lab{scalar0001}
\ee
We note that in the rotated system, we have the relation
 \be 
{\bf Q} \cdot {\bf r} =Qx. \lab{scalar0002} 
\ee
Using Eqs.~\re{scalar0001} and \re{scalar0002},
the transition amplitude \re{eq:a15} follows in the form
\begin{eqnarray}\nonumber 
a_{f,i}&=&\langle\psi_f|\left(\sigma_0-\frac{v_F}{c}\sigma_x\cos\theta\right)
e^{iQx}|\psi_i\rangle+\\ 
&+& \frac{v_F}{c}\langle\psi_f|\sigma_y\sin\theta\, e^{iQx}|\psi_i\rangle,
\label{ampl01}
\end{eqnarray}
where we can read off the relations
\begin{equation} 
Q=\sqrt{q^2+\Omega_{f,i}^{2}/c^2},\quad
\cos\theta=\frac{\Omega_{f,i}}{cQ},\quad
\sin\theta=\frac{q}{Q}.
\end{equation}
For the transition from the initial state $|\psi_i\rangle$ with energy
$E_i$ to the final state $|\psi_f\rangle$ with energy
$E_f$, the transition probability can thus be written as
\begin{equation}
|a_{f,i}|^2=\frac{q^2}{Q^2}\left(\frac{q^2}{Q^2}|F^{f,i}|^2+\frac{v_F^2}{c^2}|G_{y}^{f,i}|^2\right),
\label{prob}
\end{equation}
with the matrix elements
\begin{eqnarray}\nonumber
    F^{f,i}&=&\langle\psi_f|e^{iQx}|\psi_i\rangle-\frac{Qv_F}{\Omega_{f,i}}\langle\psi_f|\sigma_x\,e^{iQx}|\psi_i\rangle,\\
 G_{y}^{f,i}&=&\langle\psi_f|\sigma_y e^{iQx}|\psi_i\rangle.
\end{eqnarray}
Within the sudden-perturbation approximation, which approximates the narrow Gaussian pulse
in Eq.~\eqref{eq:13} by a $\delta$-function shape, Eqs.~\re{ampl01} and \re{prob} represent exact 
expressions for transition amplitudes and probabilities, respectively.
Given the initial and final state wave functions specified 
in Sec.~\ref{sec2} and in Ref.~\cite{Novikov},
one can thus compute probabilities and cross sections for different transitions, such as excitation, ionization, or pair creation. In addition, one may study   
the angular distribution of the electron emission cross section,
\be \label{crosssect}
\sigma(\vartheta)=\int_0^\infty dp\, |a_{f,i}|^2,
\ee
where $\vartheta$ is the emission angle of the electron (determined by the final state) 
and $p$ its momentum.

Finally, let us note that our assumption of a single K valley 
in graphene's band structure, see Eq.~\eqref{hamilt}, implies that the above approach can only be applied 
if the momentum transfer $q$ due to the interaction with the incident photons is small
compared to the momentum distance $k$ between K and K' valleys,
 given by $k=\frac{4\pi}{\sqrt{3}a_0}$ with
graphene's lattice constant $a_0$ \cite{CastroNeto}.
For $q\agt k$, one has to explicitly take into account both K and K' valleys, since
scattering processes connecting them become important.
For the results shown in Sec.~\ref{sec4}, we made sure that $q$ is always small against $k$.  
In that case, K valley and spin degeneracies cause only trivial overall degeneracy factors in our results.

\begin{figure}[t!]
\centering
\includegraphics[totalheight=0.24\textheight]{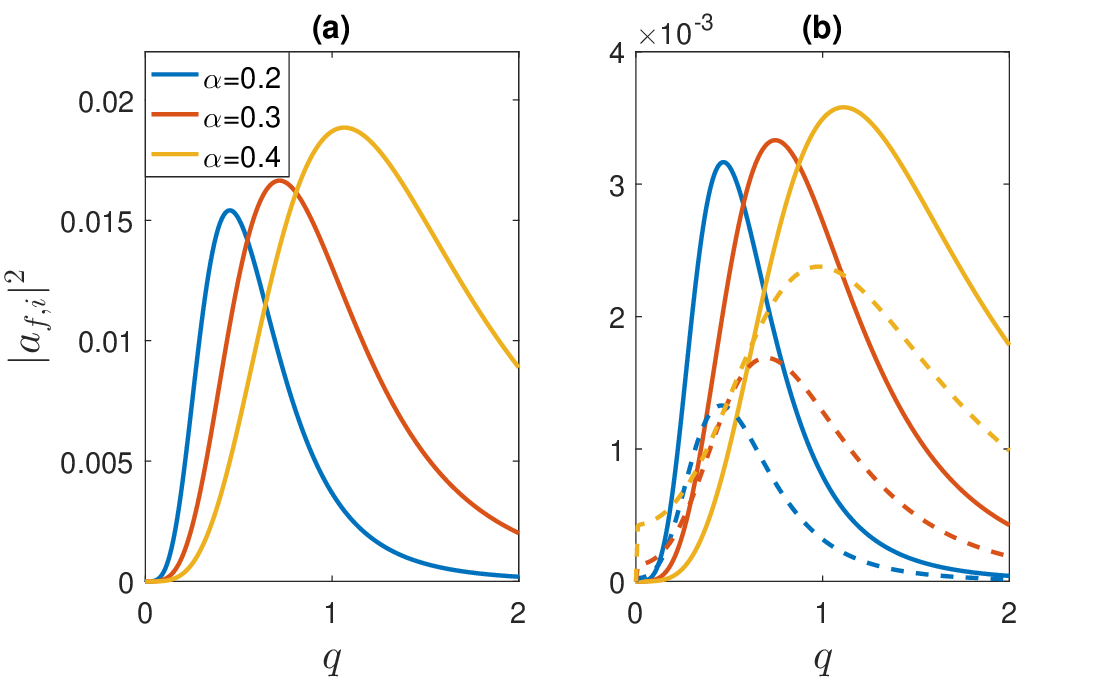}
 \caption{Excitation probability vs momentum transfer $q$ 
 from the ground state of the artificial atom to the first excited state (a), second excited state (b, solid lines) and third excited state (b, dashed lines)
 due to an ultrashort pulse, for several values of the
 Coulomb impurity charge parameter $\alpha$ in Eq.~\eqref{hamilt}. }
  \label{fig2}
\end{figure}

\section{Excitation, ionization, and pair creation processes}\label{sec4}

We next present results of the formalism in Sec.~\ref{sec3} to study
excitation, ionization, and  particle-antiparticle pair creation processes
induced by an external ultrashort electromagnetic pulse
in an artificial atom in graphene. Here, ``excitation'' is defined as the transition of an electron from 
a given initial bound state $|\psi_i\rangle$ to another bound state $|\psi_f\rangle$ in the discrete spectrum (see Sec.~\ref{sec2} and Ref.~\cite{Novikov}), 
while ``ionization'' refers to transitions from bound to continuum states described by the scattering 
state in Eq.~\eqref{scat}. In the following, we use Eq.~\re{prob} to calculate the corresponding transition probabilities.
A third process, possible only with relativistic atoms, is the particle-antiparticle pair creation, where a state 
$|\psi_i\rangle$ from the filled Dirac sea is excited to a state $|\psi_f\rangle$ in the upper continuum.
The final state is then characterized by the creation of an electron-hole pair.  The corresponding
electron and hole wave functions have also been specified in Sec.~\ref{sec2}.
A remarkable feature of the transition amplitude in Eq.~\re{ampl01} is the fact that it does not 
explicitly contain the parameters determining the electromagnetic pulse, 
such as the field amplitude $E_0$, the frequency $\omega_0$, or
the pulse duration $\lambda$.  All these parameters appear only via the momentum transfer parameter $q$
in Eq.~\re{eq:a14}.  In what follows, transition probabilities and cross sections are therefore shown 
as a function of $q$.

\subsection{Excitation probabilities}

Let us start with the excitation probabilities involving transitions $|\psi_i\rangle\to |\psi_f\rangle$ between bound states.  We assume that $|\psi_i\rangle$ is given by 
the lowest-energy state (i.e., the ground state) of the artificial atom.
 In Fig.~\ref{fig2}(a), excitation probabilities 
to the first excited state are shown for several Coulomb charge parameters $\alpha$ 
as a function of the momentum transfer parameter $q$.  
With increasing $q$, the excitation probability first increases, reaches a maximum, 
and then decreases again.  The position $q=q_{\rm max}$ of the maximum depends on $\alpha$, with larger values for $q_{\rm max}$ for larger $\alpha$.  However,
qualitatively, the curves are similar for different $\alpha$.
Next we turn to transitions to higher excited states of $s$-wave orbital type.
Figure \ref{fig2}(b) shows
the corresponding excitation probabilities from the ground state to the second 
and third excited state, respectively.  In comparison to the first excited state,  the transition
probabilities are smaller by more than one order of magnitude, and require larger momentum transfer $q$ for sizeable values.  In particular, the probabilities for the third excited state become very small.

\begin{figure}[t!]
\centering
\includegraphics[totalheight=0.24\textheight]{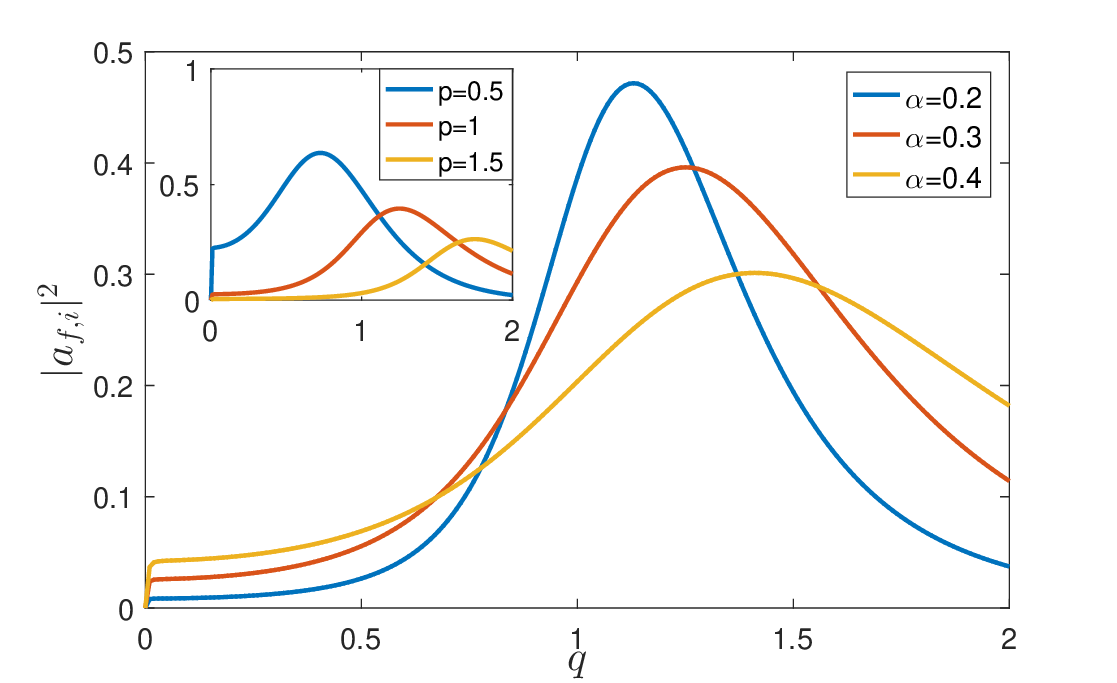}
 \caption{Ionization probability from the ground state vs momentum transfer $q$, for different Coulomb charge parameters $\alpha$ with fixed momentum $p=1$ of
 the emitted electron. The inset shows results $\alpha=0.3$ and different $p$.} \label{fig3}
\end{figure}

\subsection{Ionization probabilities}

We now turn to a discussion of the ionization probability.  Figure \ref{fig3} shows the
corresponding probability for ionization out of the ground state of the artificial atom 
as a function of the momentum transfer parameter $q$, again for different values of Coulomb charge parameter $\alpha$ and for a fixed value of the final (scattering) state momentum $p$.
Interestingly, in Fig.~\ref{fig3}, we observe a finite ionization probability already at 
small momentum transfer $q$, in contrast to the excitation probabilities in Fig.~\ref{fig2}.
Assuming a small and fixed value of $q$, the ionization probability becomes smaller with increasing 
nuclear charge parameter $\alpha$.  This observation can be rationalized by noting that the ground state will have a larger binding energy for larger $\alpha$.
In the inset of Fig.~\ref{fig3}, for a fixed value of $\alpha$, we illustrate the dependence of the
ionization probability on the scattering momentum $p$ of the final state.   
We observe that for small $q$, the ionization probability becomes smaller with increasing final-state
momentum $p$.

\begin{figure}[t!]
\centering
\includegraphics[totalheight=0.24\textheight]{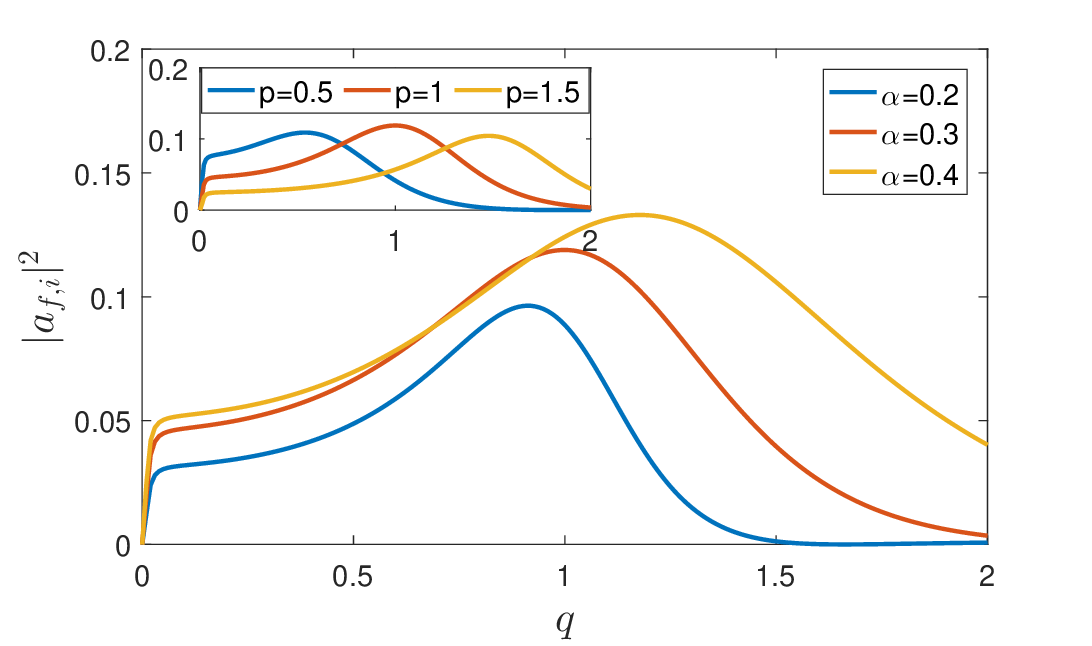}
 \caption{Probability for electron-hole pair production from the ground state due to an ultrashort pulse vs momentum transfer $q$. Results are shown for different values of $\alpha$, choosing $p=1$ for the momentum of the emitted electron. The inset shows results for $\alpha=0.3$ and different $p$. }
 \label{fig4}
\end{figure}

\subsection{Pair creation probabilities}

Next, in Fig.~\ref{fig4}, we present results for the pair production probability as a function of the
momentum transfer $q$, again for several values of $\alpha$.
For pair production induced by photon-atom (or atom-photon) collisions, one may 
consider ``free-free'' and ``bound-free" processes \cite{Eichler}.  In bound-free pair production,  
an electron is transferred from the filled Dirac sea to an atomic bound state.  We here consider 
the case of free-free pair production, where an electron is transferred from the Dirac sea
to a continuum state in the upper continuum.  
Unlike what we found for the excitation and ionization probability, for small $q$,
the pair production probability now increases with increasing nuclear charge parameter $\alpha$. Furthermore, the curves are qualitatively different and 
less sharp. In Fig.~\ref{fig4}, 
we also illustrate the corresponding dependence on the final momentum $p$ of the hole state.
The above results illustrate the dependence of transition probabilities and cross sections mainly as function of the momentum transfer parameter $q$.  It is straightforward to explore the dependence of these quantities on detailed pulse parameters, such as the strength, frequency and duration of the pulse, by using Eq.~\eqref{eq:a14}.

\section{Conclusions}\label{sec5}

In this work, we have studied electronic transitions in 2D artificial atoms in gapped graphene monolayers.  
Here an atom is created by the presence of a Coulomb impurity, and the transitions are induced by an ultrashort
electromagnetic pulse.  For this scenario, we have developed and applied a nonperturbative theoretical approach, based on a modified version of Matveev's sudden-perturbation approach \cite{VI},
for the calculation of electronic excitation, ionization, and electron-hole pair creation probabilities and cross sections.
This approach becomes exact in the limit of $\delta$-type pulses.  
The essential assumption behind our theory is therefore that the pulse is very short.

Analytical expressions for the electronic transition amplitude have been
obtained from the time-dependent Dirac equation, where we find that the dependence on external field parameters enters solely via the effective momentum transfer parameter $q$ in Eq.~\eqref{eq:a14}. Using this theoretical approach, the probabilities for excitation, ionization, and 
pair creation can then be computed as a function of the momentum transfer $q$. 
%In addition, we have also presented results for the angular dependence of the cross sections for ionization and pair creation. 

The proposed model for electronic transitions in artificial relativistic atoms in graphene could be directly
realized using available table-top experimental setups \cite{Luican,Wang,Wang2}. In that sense, 
we are confident that our results can pave the way for an experimental realization of the corresponding
ultrarelativistic atomic quantum processes in a 2D version, while their 3D counterpart still seems out
of experimental reach.  Future work may also consider setups with several Coulomb impurities, corresponding to artificial molecules or dipoles \cite{Ale2014}, in order to study the respective electronic transition probabilities.

\acknowledgements
We acknowledge funding by the Grant REP-05032022/235 (``Ultrafast phenomena and vacuum effects in relativistic artificial atoms created in graphene"), funded under the MUNIS Project, supported by the World Bank and the Government of the Republic of Uzbekistan.

\bibliography{biblio}
\end{document}